# Proliferation of anomalous symmetries in colloidal monolayers subjected to quasiperiodic light fields

Jules Mikhael [a], Michael Schmiedeberg [b], Sebastian Rausch [a], Johannes Roth [c], Holger Stark [d] & Clemens Bechinger [a,e,1]

[a] 2. Physikalisches Institut, Universität Stuttgart, Pfaffenwaldring 57, 70569 Stuttgart, Germany

[b] Department of Physics and Astronomy, University of Pennsylvania, 209 South 33rd Street, Philadelphia, Pennsylvania 19104, USA

[c] Institut für Theoretische und Angewandte Physik, Universität Stuttgart, Pfaffenwaldring 57, 70569 Stuttgart, Germany

[d] Institut für Theoretische Physik, Technische Universität Berlin, Hardenbergstraße 36, 10623 Berlin, Germany

[e] Max-Planck-Institut für Metallforschung, Heisenbergstraße 3, 70569 Stuttgart, Germany

[1] To whom correspondence should be addressed. E-mail: c.bechinger@physik.uni-stuttgart.de, Tel. +49(0)711/685-65218, Fax. +49(0)711/685-65285.

**Quasicrystals provide a fascinating class of materials with intriguing properties. Despite a strong potential for numerous technical applications, the conditions under which quasicrystals form are still poorly understood. Currently, it is not clear why most quasicrystals hold 5- or 10-fold symmetry but no single example with 7 or 9-fold symmetry has ever been observed. Here we report on geometrical constraints which impede the formation of quasicrystals with certain symmetries in a colloidal model system. Experimentally, colloidal quasicrystals are created by subjecting micron-sized particles to two-dimensional quasiperiodic potential landscapes created by n=5 or seven laser beams. Our results clearly demonstrate that quasicrystalline order is much easier established for n = 5 compared to n = 7. With increasing laser intensity we observe that the colloids first adopt quasiperiodic order at local areas which then laterally grow until an extended quasicrystalline layer forms. As nucleation sites where quasiperiodicity originates, we identify highly symmetric motifs in the laser pattern. We find that their density strongly varies with *n* and surprisingly is smallest exactly for those quasicrystalline symmetries which have never been observed in atomic systems. Since such high symmetry motifs also exist in atomic quasicrystals where they act as preferential adsorption sites, this suggests that it is indeed the deficiency of such motifs which accounts for the absence of materials with e.g. 7-fold symmetry.**

The presence or lack of order is of primary importance in a broad range of fundamental phenomena in science. Until the early 1980s, it was unanimously established that ordered matter is always periodic (1). Accordingly, the rotational symmetry in real space was thought to be limited to N=2,3,4 and 6. However some metal alloys (2), polymers (3), micelles (4), and even recently colloidal nanoparticles (5) and non-spherical particles (6), have defied these crystallographic rules and self-organized into so-called quasicrystals. These structures form a novel type of matter which - unlike periodic crystals or amorphous materials - exhibit long-range positional order but are not periodic. Quasicrystals show many interesting properties which are quite different compared to that of periodic crystals. Accordingly, they are considered as materials with high technological potential e.g. as surface coatings, thermal barriers, catalysts or photonic materials (7).

Since the properties of quasicrystals are strongly connected to their atomic structure, a better understanding of their growth mechanisms is of great importance(8-11). Perhaps one of the most interesting questions in this context is why all observed quasicrystals have only 5-, 8-, 10-, and 12-fold symmetry but no single quasicrystal with 7-, 9-,11-, and 13-fold symmetry was ever found (12). For a classification of different surface symmetries it is helpful to consider the rank $D$, i.e. the number of incommensurate wave vectors required to define the reciprocal lattice of a $d$-dimensional structure (13). For two dimensional lattices with N-fold symmetry, D is the number of positive integers less than and relatively prime to N (14). A table listing the values of D for $2 \leq N \leq 12$ is included in the supplementary table 1. For $D = d$ the structure is periodic, for $D > d$ it becomes quasiperiodic. Interestingly, the diffraction patterns of all experimentally observed two-dimensional quasicrystals fulfill $D = 4$. In contrast, structures with $D = 6$ (e.g. 7-fold symmetry) or with $D > 6$ are experimentally not observed. The goal of our work is to understand why the formation of solids with $D \geq 6$ is hindered or even prohibited compared to $D = 4$ since this will give important insights into the fundamental principles on how ordered matter arises in general (12, 15, 16).

The peculiarity of quasicrystals with rank D=6, for instance, was lately illustrated by simple geometrical arguments (Fig.1) (17). Starting with a pentagonal, hexagonal or heptagonal tile, one can construct 5-, 6- or 7-fold structures by iteratively adding identical clusters at the free edges.

Although the building complexity of all three structures is quite similar, their decoration with atoms or molecules is not. Contrary to the 5-fold structure which leaves voids (white), the 7-fold structure causes overlaps of different shapes (green). Their atomic decoration becomes nontrivial due to steric hindrance. Similar, non-uniform overlaps, however, also occur for e.g. 8-fold structures (being indeed observed in atomic systems); therefore additional reasons for the absence of 7-fold quasicrystals must exist.

**Experimental Procedure**

Because regular or semiregular polyhedra with more than 5-fold symmetry do not exist, any quasicrystal with N>6 will be comprised of periodically stacked quasicrystalline layers (18). In particular, this argument would also apply to a hypothetical quasicrystal with 7-fold symmetry. Therefore, in our work we are searching for hurdles that impede the formation of certain symmetries in two-dimensional systems. To induce quasiperiodic order of different symmetries – even such not found in nature – we apply an external quasiperiodic potential landscape to a system of micron-sized colloidal particles and study how the latter assumes quasiperiodic order when the potential strength is increased. Despite their much simpler pair potential, colloids show many striking similarities with atomic systems and can therefore be regarded as mesoscopic model systems (19). As a consequence of their length scale, they offer the unique opportunity to investigate this process in real space and real time.

Experimentally, we created quasiperiodic potentials with rank D=4 and rank D=6 by interfering n=5. and 7 laser beams with parallel polarization (20, 21). The resulting intensity patterns have 10- and 14-fold symmetry and posses a characteristic length scale $a_v$ determined by the projection of the reciprocal wave vectors onto the sample plane $G_j = 2\pi / a_v$ (22). The light patterns exert optical forces (23) onto colloidal particles and can thus provide an energy landscape with quasiperiodic order (20). As a particular advantage compared to atomic systems, the potential strength can be continuously varied by the laser intensity $I_0$. In our study, we have used negatively charged polystyrene particles dispersed in water with radius R=1.45 μm. Their pair potential is given by u(r) ~ exp(−κr)/r with r the particle distance and $\kappa^{-1}$ the Debye screening length being set by the ion concentration of the solvent. The colloidal suspension is contained in a flat silica cuvette where they form a two-dimensional system. The particle positions and the quasiperiodic interference patterns were simultaneously monitored with an inverted microscope and two CCD cameras which allowed us to obtain the particle positions relative to the quasiperiodic light patterns with a spatial resolution of 150 nm.

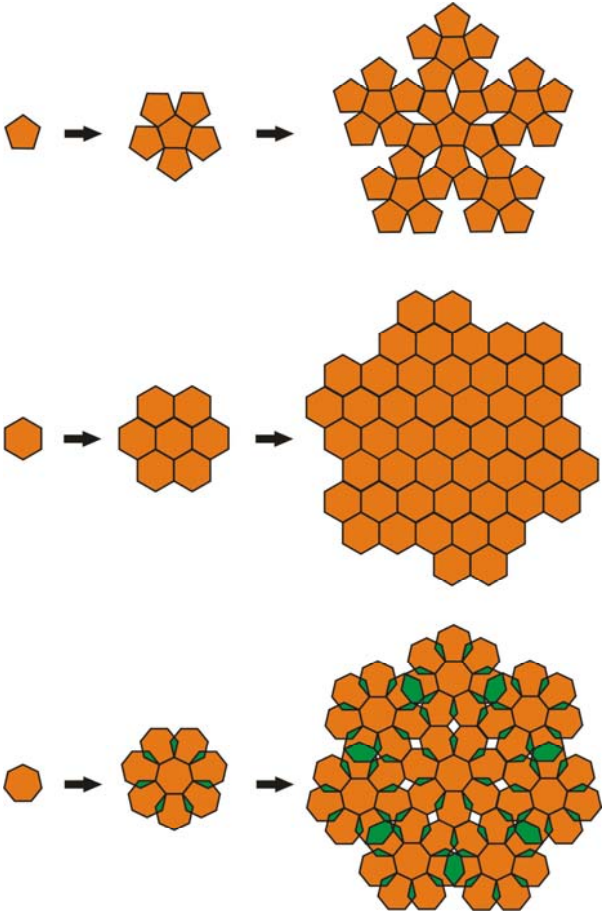

*Figure 1: Construction of 5-, 6- or 7-fold tilings.*
*Assembly of periodic and quasiperiodic tilings according to the method of polysynthetic twinning (17). The construction principle consists in starting with a nucleus polygon and then iteratively adding others to the free edges. Pentagons lead to a tiling with quasiperiodic 5-fold local symmetry including gaps (white). From hexagons a periodic tiling without any gaps is obtained. For heptagons, the tiling contains both gaps (white) and non-uniform overlapping regions (green) leading to non-trivial steric hindrance problems when decorating the tiles with atoms.*

**Results and Discussion**

Figs.2a,e shows the measured intensity distribution of quasiperiodic light patterns created from 5 and 7 laser beams which exhibit differently sized pentagonal and heptagonal structures. The characteristic length scales of both patterns has been adjusted to the same value of $a_v = 7.5$ µm. In figs. 2b,c,f,g,h we qualitatively compare how a colloidal system responds to these patterns as a function of the areal laser intensity $I_0$. The particle density and the Debye screening length were chosen such, that the colloids form a liquid in the absence of the laser field. In case of the 5-beam pattern, the colloids follow the underlying potential already at quite small intensities as seen by the formation of local pentagons. At a laser intensity of about $I_0 = 2.82$ µW µm$^{-2}$ the particle configuration almost perfectly resembles the underlying interference pattern. In contrast, when exposing the colloidal system to a 7-beam pattern, the particles do only rarely form heptagons at the same intensity. Even at $I_0 = 5.08$ µW.cm$^{-2}$ (fig.2h) the particles do not equally respond to the pattern compared to fig.2c. To quantify this observation, in fig.2d we show the corresponding bond orientational order parameter (see Materials and Methods) which measures the alignment of the particles along the 5 and 7 directions of the laser fields (see dashed lines in Figs.2a,e). It is clearly seen that at 2.82 µW µm$^{-2}$ the particles on the 5 beam interference pattern have achieved almost 70% of the saturation value while in case of the 7 beam interference pattern even for the highest experimental intensity $I_0 = 5.08$ µW µm$^{-2}$ only 40% of the corresponding maximum value were achieved. Similar results were obtained for other particle densities and Debye screening lengths. Obviously, quasicrystalline order induced by a quasiperiodic light potential of $n = 7$ is only achieved at much higher potential strengths of the light field compared to $n = 5$.

To understand on a microscopic scale what prevents the colloidal particles to follow the 7 beam quasiperiodic pattern at small laser intensities, we increased their density and electrostatic interaction which results in the formation of large hexagonal colloidal domains in the absence of the laser field. Due to the well defined particle positions in a periodic crystal (compared to a liquid) this facilitates the analysis how the particles respond to a quasiperiodic potential landscape. Figs.3a-c show typical particle snapshots showing how a colloidal crystal responds to a 7-beam pattern of increasing laser intensity. In order to highlight regions where the hexagonal order becomes destroyed due to the interaction of the particles with the quasiperiodic light field, we have applied a Delaunay triangulation (lines) and encoded the number of nearest neighbors with different colors (green=5, blue=6, red=7). With this representation one can easily identify regions where the initial hexagonal structure of the particles becomes distorted by the underlying laser field. For $I_0=0$ the particles arrange in a single hexagonal domain with most of the particles having 6 nearest neighbors. Upon increasing $I_0$, the interaction with the quasiperiodic laser field leads to an increasing number of defects. Interestingly, these defects develop at rather localized regions (center of Fig.3a) and remain there during the entire measurement. Further increase of $I_0$ leads to a spatial extension of the defect area (Fig.3b) until finally most of the particles lost their 6-fold coordination (Fig.3c). In this situation, the particles almost perfectly follow the underlying quasiperiodic light field as seen by the 14-fold rotational symmetry of the corresponding structure factor (Fig.3d). To identify what determines the positions where the defects first occur, in Figs.3e-g we superimposed the particle positions taken from Figs.3a-c with the laser intensity distribution (light grey). In red we marked those colloids which most strongly respond to the quasiperiodic laser lattice, i.e. whose nearest neighbors lost their original 6-fold coordination and which overlap with quasiperiodic potential wells (all other particles are white). As seen from Fig.3e, the particles which most strongly respond to the interference pattern are those at so-called high symmetry stars (yellow) which correspond to local motifs in the laser lattice having the highest possible local rotational symmetry. High symmetry stars are comprised of a central potential well surrounded by 14 others. Since the potential wells are rather deep, they enforce a colloidal arrangement with quasiperiodic order. Upon further increasing $I_0$, quasicrystalline order laterally spreads around the high symmetry stars until almost all particles lost their 6-fold coordination and follow the quasicrystalline potential (Fig.3h). From this we conclude, that the high symmetry stars of the laser pattern act as preferential adsorption sites where quasicrystalline order is initiated in the colloidal monolayer. Accordingly, the density of these stars should largely determine how easily quasicrystalline order will laterally proliferate across the system. Interestingly, the particles do not equally respond to the heptagonal motifs (green circles in Figs. e-g) of the laser pattern although their potential depths is similar to that of the high-symmetry stars. This may be caused by the additional central potential well in a high symmetry star which stabilizes the decoration of high symmetry stars with colloidal particles.

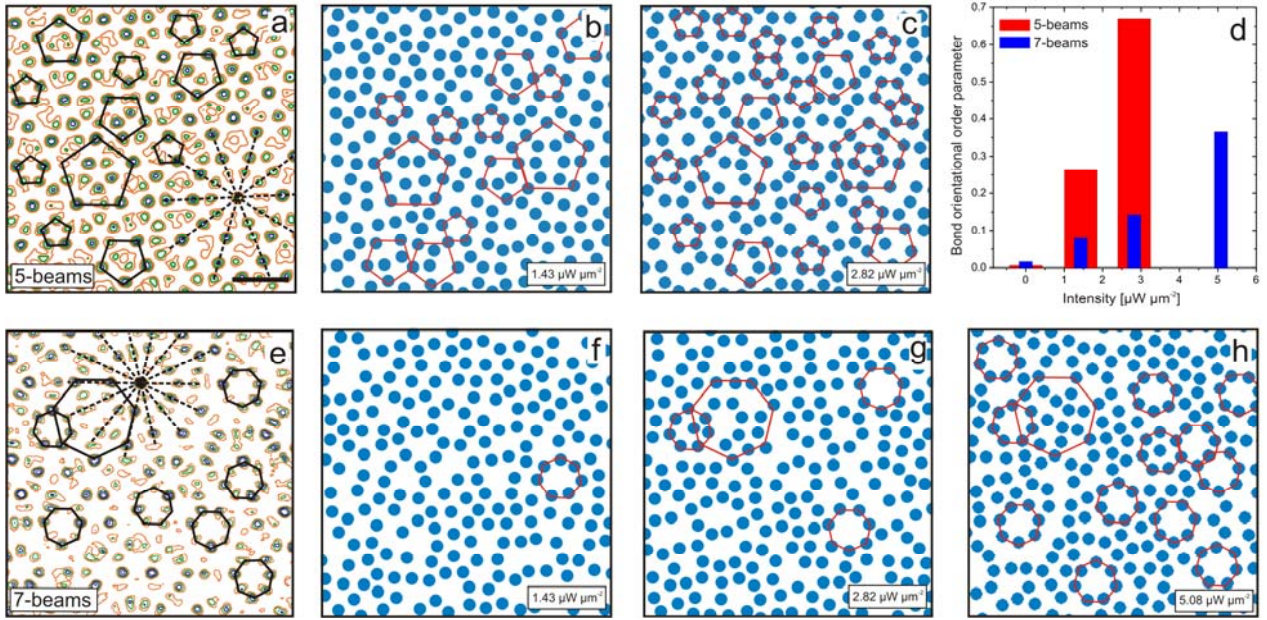

*Figure 2: Colloidal monolayer subjected to 5 and 7-beam interference patterns.*
*a,e Contour plots of the intensity distribution of a 5-beam (a) and a 7-beam (e) pattern. They exhibit characteristic motifs (shown as black solid lines), pentagonal in (a) and heptagonal in (e). The dashed lines correspond to the directions of the incident laser beams. The scale bare denotes 15 μm. b,c,f,g,h Comparison of typical configurations of colloidal particles in the presence of the 5-beam (b,c) or the 7-beam (f-h) pattern for different laser intensities ($\phi = 0.028$ μm$^{-2}$, $\kappa^{-1} \approx 100$nm). d, Bond-orientational order parameter of the colloidal system on quasiperiodic patterns of 5 and 7 laser beams.*

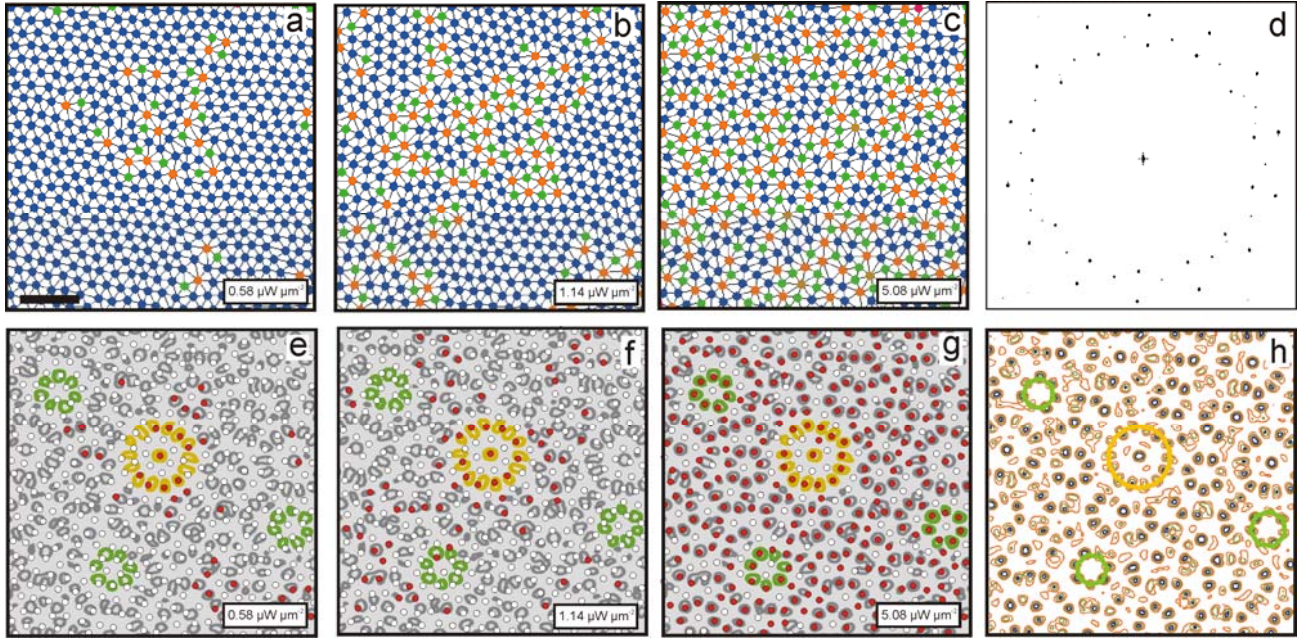

*Figure 3: Phase transition from a periodic crystal to a quasicrystal on a 7-beam pattern.*
*a-c, Delaunay triangulation of typical particles configurations for a colloidal monolayer interacting with a 7-beam interference pattern of increasing intensity: $\phi = 0.033$ μm$^{-2}$, $\kappa^{-1} \approx 200$nm, $a_v = 8.5$ μm. The particle coordination 5, 6, 7 is encoded in green, blue and red, respectively. The scale bar corresponds to 20 μm. d, Two-dimensional structure factor calculated for the particle configuration in c. e-g, Particle positions (taken from a-c) superimposed to the light pattern (grey contour). Particles, whose nearest neighbors are not 6-fold coordinated, and overlap by more than 50% with an intensity maximum of the laser field are colored in red (white otherwise). The yellow and green patterns indicate distinct symmetric motifs in the laser pattern and are a guide to the eye. h, Contour plot of the 7-beam light pattern with heptagons (green) and high symmetry stars (yellow).*

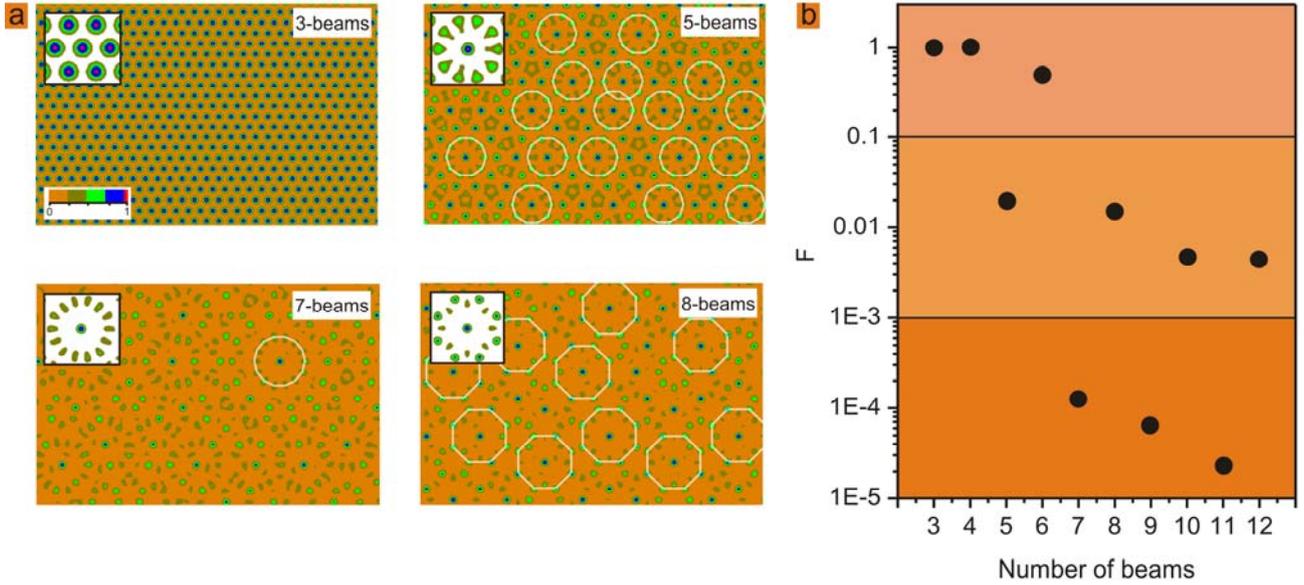

*Figure 4: High-symmetry star densities*
*a, Calculated 3-, 5-, 7- and 8-beam interference patterns. The white polygons highlight high symmetry stars (note that in the 3-beam pattern, every well is the center of a high symmetry star). The inset shows a single high symmetry star for each pattern. b, Number density of high symmetry stars F calculated for n-beam patterns with $3 \leq n \leq 12$. The density of high symmetry stars is defined by the fraction of wells deeper than 99% of the deepest well. For 11-beam patterns the value provides an upper limit.*

High symmetry stars occur in any interference pattern created with n beams. Fig.4a illustrates such patterns of n=3, 5, 7, 8 laser beams together with the corresponding high symmetry stars. Depending on whether *n* is even or odd, they exhibit either *n*- or 2*n*-fold rotational symmetry. Fig.4b shows the calculated number density *F* of such stars for interference patterns created with $3 \leq n \leq 12$ laser beams. In our calculations, high symmetry stars have been identified by an appropriate intensity threshold in the light field (for details see Materials and Methods). Although F varies by several orders of magnitude depending on n, three regimes can be distinguished: for periodic laser patterns with $n=3$, 4 or 6 laser beams, $F$ is of the order of 1 (i). For $n=5$, 8, 10 and 12, i.e. for laser fields with rank $D=4$, the value of $F$ is around a factor of 100 smaller (ii) and for $n=7$, 9, 11 ($D \geq 6$) the density of high symmetry stars further decreases significantly (iii). In particular the large difference in F for n=5 and 7 now explains why colloidal quasiperiodic order was rather difficult to achieve in laser patterns created with 7 laser beams compared to 5. It should be mentioned, that the observed F-dependence is a rather generic property of quasiperiodic structures which is not limited to interference patterns but also found in $n$-fold rhombic tilings (see supplementary figure S1).

It is important to realize that all quasicrystalline symmetries in regime (ii) have been experimentally observed in atomic systems (17) while no single example of regime (iii) was ever found. This remarkable coincidence cannot be accidental. It suggests that a similar argument, i.e. a deficiency of high symmetry motifs, may also explain why certain quasicrystalline symmetries are not observed in atomic systems. In contrast to crystals which are periodic in all three dimensions, quasiperiodicity is always (except for icosahedral quasicrystals) restricted to two dimensions (24, 25). Accordingly, three-dimensional quasicrystals are comprised of a periodic stacking of quasiperiodic layers and any hurdle in the formation of quasiperiodic order within a single layer will eventually prohibit their growth along the periodic direction. Indeed, strong evidence for the importance of symmetric motifs for the formation of quasiperiodic order in atomic systems has been recently obtained by scanning-tunneling microscopy studies and ab initio calculations (26, 27) where the structure of thin overlayers adsorbed on quasiperiodic surfaces was determined. Only for certain atoms, which do preferentially absorb on these motifs, extended thin films with quasiperiodic order have been observed.

**Conclusion**

From our experiments we conclude that colloidal particles on quasicrystalline laser fields favor symmetries with rank D=4 compared to D=6 in agreement with atomic systems. We have shown that this is attributed to large differences in the number density of highly symmetric sites where quasicrystalline order first originates. The importance of geometrical considerations for the formation of quasicrystals is also in agreement with recent simulations which showed that dense packings of entropically interacting tetrahedra can lead to dodecagonal quasicrystals (6). Structural properties play also an important role for the friction (28), lattice vibrations (29) and electronic properties (30) of atomic systems with quasiperiodic long-range order. Because of this strong link between the structure and the physical behaviour, we currently explore the possibility to create free-standing quasicrystalline colloidal structures with unusual rotational symmetries with a view to potential applications as photonic or phononic devices.

**Materials and Methods**

As colloidal particles we used polystyrene particles with a radius of R=1.45μm, polydispersity of 4% and a negative surface charge density of 9.8μC/cm$^2$ (IDC, Batch#1212). The particle density is adjusted by means of an optical fence was created by a laser beam (λ=488nm) being scanned around the central region of the sample (20). An additional laser beam (λ=514nm) which is vertically incident on the sample cell from above pushes the colloids towards the negatively charged substrate and reduces vertical particle fluctuations to less than 5% of the particle radius. Therefore the system can be considered as two-dimensional. The interference patterns were created with a 18W frequency-doubled solid state laser (Verdi V-18, Coherent) which was splitted into 5 and 7 beams which were overlapped in the sample plane. The characteristic length scales $a_v$ of the interference patterns was varied by controlling the distance between the beams in front of an achromatic lens which focuses them in the sample plane.

Due to the absence of periodicity, a translational order parameter is not suitable to describe quasiperiodic order. Instead we used the m-bond orientational order parameter (22)

$$\psi(m) = \left\langle \left| \frac{1}{N} \sum_{j=1}^{N} \frac{1}{n_j} \sum_{k=1}^{n_j} e^{im\Theta_{jk}} \right| \right\rangle.$$

Here the inner sum is over all $n_j$ nearest neighbors of particle j, N the total number of particles, and $\Theta_{jk}$ the angle of the bond connecting particles j and k measured with respect to an arbitrary reference direction. ψ(m) is non-zero, if the bonds towards a nearest neighbor particle are mainly oriented along m directions, i.e., for colloids in a strong laser field created by 5 beams, ψ(10) is non-zero because there are 10 possible bond-directions. Similarly, in the 7-beam pattern ψ(14) is non-zero. The bond orientational order parameter in fig. 2d is normalized such that it reaches a value of one if all colloids occupy the deepest available potential wells.

To quantify the number density of high symmetry stars F we calculated the potentials that correspond to the intensity distribution of the laser fields (see also (31), (32)) and determined the fraction of local minima whose depths is at least 99% of the deepest well. At a deep potential well the laser beams are almost in phase and therefore such a minimum is the center of a high symmetry star.

**Acknowledgements**

We acknowledge Laurent Helden for the technical support. This work is financially supported by the Deutsche Forschungsgemeinschaft.

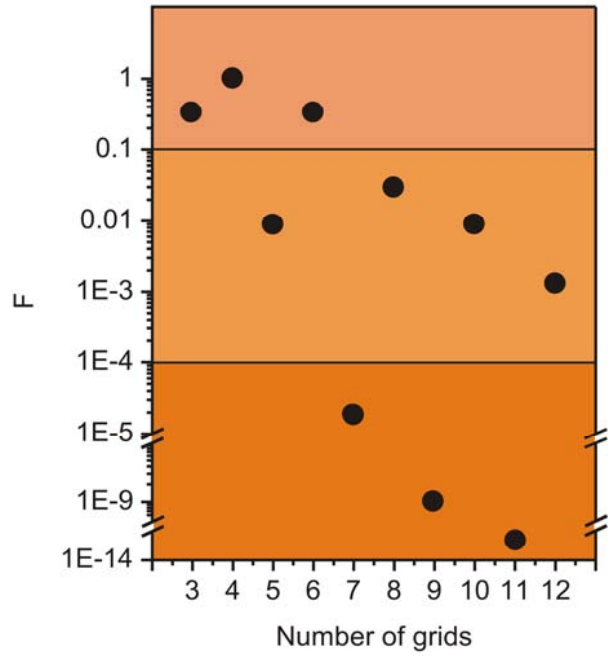

*Fig. S1 Number density of high symmetry stars F calculated for rhombic tilings. These tilings are constructed from n grids of parallel lines pointing along n symmetrically distributed directions and have the same rotational symmetry as n-beam light pattern*

| N | 2 | 3 | 4 | 5 | 6 | 7 | 8 | 9 | 10 | 11 | 12 |
|---|---|---|---|---|---|---|---|---|---|---|---|
| D=Φ(N) | 1 | 2 | 2 | 4 | 2 | 6 | 4 | 6 | 4 | 10 | 4 |

*Table S1: Euler totient function Φ(N) calculated for lattices with N-fold real space symmetry ($2 \leq N \leq 12$). For two-dimensional lattices, the rank D is related to the N-fold rotational symmetry by Φ(N)*